\documentclass[useAMS, usenatbib]{mn2e}
\usepackage{epsfig}
\usepackage{graphicx}

\newcommand{\FIGwindow}{
  \begin{figure}
    \includegraphics[width=78mm]{fig1.window.epsi}
    \caption{  \label{fig:windz}
      Window functions of current (solid) and next-generation (solid, bold)
      \xray\ surveys are shown along with the cumulative fraction,
      $N(<z)/N_{tot}$, (dashed) of {\emph all} halos at \mhiscale\ (upper)
      and \mloscale\ (lower). 
      } 
  \end{figure}
}

\newcommand{\FIGlmanalytic}{
  \begin{figure}
     \centering
    \includegraphics[height=60mm]{fig2.lmanalytic.epsi}
  \caption{Geometric mean, $\langle L\rangle$ $\equiv e^{\langle \lpar
      \rangle}$, of the de-evolved luminosity for current (dotted) and
    next-generation (solid) flux-limited samples.  The upper panel shows
    the bias relative to the underlying, mass-limited population (dashed).
  \label{fig:lmana}}
\end{figure}}

\newcommand{\FIGltranalyticsim}{
\begin{figure}

\centering
 \includegraphics[width=80mm]{fig3.ltr_anasim.epsi}
  \caption{
  The \xray\ luminosity--temperature relation 
  (upper panels) and its dispersion $\sigl(T)$ (lower panels) expected
  for current, flux-limited ($3\times 10^{-12}\ \cgsflux$) samples and
  for different correlation coefficients, $r = \left( -0.7, 0.0, 0.7
  \right)$.  Solid lines show the analytic expectations, points
  show discrete realizations from Hubble Volume full-sky halo
  realizations, and dashed lines are power-law fits to these discrete
  samples.   
  \label{fig:ltranasim}}
\end{figure}}

\newcommand{\FIGscatterevolution}{
\hspace{1in}
\begin{figure}
\centering
\includegraphics[width=84mm]{fig4.lz_nz.T4.epsi}
\caption{Redshift behavior of the geometric mean luminosity (upper),
  and the sky surface density (lower) of $4\,\kev$ clusters above a
  next-generation flux limit ($10^{-14}\ \cgsflux$).  The solid 
  line in all panels is our default model ($s=1$, $\sigl=0.59$,
  $\lnlnorm=1.81$, $r=0$). The left panels vary the correlation
  coefficient, $r= 0.7$ (short-dashed), and $r=-0.7$ (long-dashed),
  otherwise retaining the default model parameters.  The right panels
  hold $r=0$ and show the effects of changing the evolution and
  scatter: $s=0.95$,  $\sigl=0.72$, $\lnlnorm=1.65$ (long-dashed);
  and $s=1.1$,  $\sigl=0.35$, $\lnlnorm=1.95$ (short-dashed).     
  \label{fig:ltranaz}}
\end{figure}}

\newcommand{\mpar}{\mu}
\newcommand{\lpar}{\ell}
\newcommand{\probdist}{{\cal P}}
\newcommand{\bigL}{{\cal L}}

\newcommand{\bdm}{\begin{displaymath}} 
\newcommand{\edm}{\end{displaymath}}
\newcommand{\beq}{\begin{equation}} 
\newcommand{\eeq}{\end{equation}} 
\newcommand{\beqnarr}{\begin{eqnarray}}
\newcommand{\eeqnarr}{\end{eqnarray}}
\newcommand{\bit}{\begin{itemize}} 
\newcommand{\eit}{\end{itemize}} 
\newcommand{\ben}{\begin{enumerate}} 
\newcommand{\een}{\end{enumerate}}
\newcommand{\bfi}{\begin{figure}[htb]} 
\newcommand{\bpfi}{\begin{figure}[p]}
\newcommand{\barr}{\begin{array}}
\newcommand{\earr}{\end{array}}
\newcommand{\bec}{\begin{center}}
\newcommand{\eec}{\end{center}}

\newcommand{\lta}{\sim}
\newcommand{\gta}{\gtrsim}

\newcommand{\aqual}{&=&}

\newcommand\rmd{{\rm d}}

\newcommand{\kmsmpc}{{\rm km}\,{\rm s}^{-1}\,{\rm Mpc}^{-1}}

\newcommand{\kev}{\>{\rm keV}}

\newcommand{\xray}{\hbox{X-ray}}

\newcommand\mhiscale{$10^{15}\, {h}^{-1}{\rm M}_{\odot}$}
\newcommand\mloscale{$10^{14}\, {h}^{-1}{\rm M}_{\odot}$}

\newcommand\lumunits{  10^{44}\,          {\rm ergs}\,{\rm s^{-1}} } 
\newcommand\massunits{ 10^{15}\, h^{-1}\, {\rm M}_{\odot}          }
\newcommand\fluxunits{{\rm ergs}\, {\rm s^{-1}}\, {\rm cm^{-2}}}
\newcommand\fluxref{3 \times 10^{-12}}
\newcommand\fluxng{1 \times 10^{-14}}

\newcommand{\lnlnorm}{\hbox{${\ln}L_{15,0}$}}
\newcommand{\lnM}{\hbox{${{\ln}M}$}}
\newcommand{\lnL}{\hbox{${{\ln}L}$}}
\newcommand{\lnT}{\hbox{${{\ln}T}$}}
\newcommand{\lgT}{\hbox{${{\rm log}\,kT}$}}

\newcommand{\lm}{L--M}
\newcommand{\lt}{L--T}

\newcommand\Lobs{L_{\rm obs}}
\newcommand\Lrest{L_{\rm rest}}

\newcommand\Lone{\bigL^{(1)}}
\newcommand{\Lsecondmom}{\hbox{${\bigL^{(2)}}$}}

\newcommand{\sigate}{\hbox{$\sigma_8$}}

\newcommand\sigl{\sigma_{\lpar}}

\newcommand{\ldiff}{\psi_{min}}
\newcommand{\tdist}{{\rm e}^{-\delta_t^2/2}}

\newcommand{\ltrprox}{\hat{\lpar}}
\newcommand\fluxlim{f}
\newcommand\lmin{\lpar_{min}}
\newcommand\phit{\phi_{min}}

\newcommand{\lbar}{\bar{\lpar}}

\newcommand{\ka}{-0.209}
\newcommand{\kb}{1.18}
\newcommand{\kc}{-0.39}
\newcommand{\kd}{-0.098}
\newcommand{\ke}{-0.092}
\newcommand{\kf}{+0.085}

\def \ie        {\hbox{\it i.e.,} }

\def \msol      {{\rm\ M}_\odot}

\def \hinv      {\hbox{$\, h^{-1}$} }

\def \keV       {\hbox{$\,\kev$}}

\def \cgsflux   {{\rm\ erg\ s^{-1}\ cm^{-2}}}

\def \se        {\!=\!}
\def \sims      {\sim \!}

\def \ssimeq    {\! \simeq \!}

\def \spropto   {\! \propto \!}

\def\myputfigure#1#2#3#4#5% {\hskip0.03\textwidth\vskip#5pt%\hfill
{\hskip0.03\textwidth\vskip#5pt%\hfill
\makebox[0pt]{\hskip#2in
\includegraphics[width=#3\textwidth]{#1}}\vskip#4pt\hfill} % from Andy Pawl
\def\spose#1{\hbox to 0pt{#1\hss}}
\def\lta{\mathrel{\spose{\lower 3pt\hbox{$\mathchar"218$}}
     \raise 2.0pt\hbox{$\mathchar"13C$}}}
\def\gta{\mathrel{\spose{\lower 3pt\hbox{$\mathchar"218$}}
     \raise 2.0pt\hbox{$\mathchar"13E$}}}

\begin{document}
\title[Effects of Selection and Covariance on X--ray Scaling Relations
of Galaxy Clusters] 
  {Selection, Covariance, and Scaling Relations of Galaxy Clusters}
\author[B. Nord et al.]
  {B.~Nord,$^1$\thanks{Email: bnord@umich.edu}
  R.~Stanek,$^2$ 
  E.~Rasia,$^1$ $^4$
  A.E.~Evrard,$^1$ $^2$ $^3$ \\
  $^1$Department of Physics, University of Michigan, 450 Church St.,
  Ann Arbor, MI  48109-1040\\ 
  $^2$Department of Astronomy, University of Michigan, 500 Church St.,
  Ann Arbor, MI  48109 \\
  $^3$Michigan Center for Theoretical Physics, University of
  Michigan, 500 Church St., Ann Arbor, MI  48109 \\
  $^4$ Chandra Fellow}
\date{Draft 1: March, 2007}

%\pagerange{\pageref{firstpage}--\pageref{lastpage}} \pubyear{2007}

\def\LaTeX{L\kern-.36em\raise.3ex\hbox{a}\kern-.15em
    T\kern-.1667em\lower.7ex\hbox{E}\kern-.125emX}

\maketitle

%%%%%%%%%%%%%%%
% Abstract    %
%%%%%%%%%%%%%%%
\begin{abstract}
We explore how the behavior of galaxy cluster scaling relations
are affected by flux-limited selection biases and intrinsic covariance
among observable properties.  Our models presume log-normal
covariance between luminosity (L) and 
temperature (T) at fixed mass (M), centered on evolving, power-law
mean relations as a function of host halo mass.  Selection can mimic evolution; 
the \lm\ and \lt\ relations from shallow X-ray flux-limited samples will 
deviate from mass-limited expectations at nearly all scales while the
relations from deep surveys ($10^{-14} \cgsflux$) become complete, and 
therefore unbiased, at masses above $\sims 2  
\times 10^{14} \hinv \msol$.  We derive expressions for low-order moments of
the luminosity distribution at fixed temperature, and show that the
slope and scatter of the \lt\ relation observed in flux-limited 
samples is sensitive to the assumed \lt\ correlation coefficient.  In
addition, \lt\ covariance affects the redshift behavior of halo counts
and mean luminosity in a manner that is nearly degenerate with intrinsic population
evolution.  
\end{abstract}


%%%%%%%%%%%%%%%%%%%%%%%%%%%%%%%%%%%%%%%%%%%%%%%%%%%%%%%%%%%%%%%%%%%%%
\section{Introduction}\label{sec:intro} %%%%%%%%%%%%%%%%%%%%%%%%%%%
%%%%%%%%%%%%%%%%%%%%%%%%%%%%%%%%%%%%%%%%%%%%%%%%%%%%%%%%%%%%%%%%%%%%%
Encoded within observed galaxy cluster populations are important
keys to cosmic structure evolution.  The decryption of these
clues requires an accurate understanding of survey selection.  The
character of the cluster population selected on an observable, ${\cal O}$, 
will depend strongly on how ${\cal O}$ relates to the underlying halo
mass and how that relation varies over time. 

Upcoming surveys at sub-millimeter and optical wavelengths will complement
existing efforts in the X-ray \citep{ebeling:98, bohringer:02, 
bohringer:06, pierre:06} and optical \citep{miller:05,koestercat:07} bands.
Survey cross-calibration will enable detailed characterizations of the
inter-relationships among observable tracers of mass.  Precision
cosmology derived from number counts and clustering
\citep{levine:02,majumbdarMohr:04,lima:04,lima:05} relies on this type
of interwoven tapestry.      

Although efforts to understand cluster covariance will ultimately be
pan-chromatic, this paper focuses on the behavior of two bulk \xray\
properties---luminosity, $L$, and temperature, $T$---derived from
observations of flux-limited samples. In particular, we investigate
how the form and evolution of the \lt\ relation depend on assumptions
regarding the covariance in these observables with mass.  

Dimensional arguments suggest power-law behaviors for the mean
luminosity and temperature as a function of mass and epoch 
\citep{kaiser:86}.  Recent work has moved beyond the slopes and
intercept of this relation, focusing on a stochastic model with
log-normal scatter about the mean population behavior
\citep{reiprich:02, stanek:06, reiprich:06}.  The physical origin of 
this scatter has been linked to non-adiabatic mechanisms, 
such as cool cores \citep{fabian:96,markevitch:98,arnaud:99,ohara:06} and
mergers \citep{rowley:04, poole:07}.  Analytic investigations discuss
the impact of entropy floors on the scatter and 
evolution of the scaling relations among luminosity,
temperature, and mass 
\citep{bower:97, balogh:06}.

Self-similar evolution of the type envisaged by \cite{kaiser:86} has
been tested with several 
observational samples, including those drawn from surveys from ASCA
\citep{mushotzky:97, novicki:02}, XMM-Newton
\citep{vikhlinin:02,kotov:05}, CHANDRA \citep{lumb:04}, and ROSAT
\citep{fairley:00,maughan:06}.  These studies compare the \lt\,
relation among redshift sub-populations in an effort to find the 
scaling with epoch.  Results range from somewhat stronger than
self-similar scaling to slightly negative.  The work
presented here addresses a potential source of confusion for such
studies.  Shallow flux-limited surveys are mass-incomplete, while
samples derived from deeper flux-limited samples become 
  complete at sufficiently high masses. This difference in selection,
essentially a redshift-dependent Malmquist bias, manifests effects
similar to genuine population evolution.    

We work with two canonical flux limits, 
$\fluxref$ and $\fluxng\,\fluxunits$, representative of the existing
REFLEX survey and of future surveys such as ChaMP, XMM-LS and
Spectrum X-Gamma\footnote{http://hea-www.harvard.edu/SXG/sxg.shtml}.
We follow the conventions used in  
\cite{stanek:06}, where $L$ is a rest-frame, soft-energy band ($0.1-2.4\, 
\kev$) \xray\ luminosity in units of $\lumunits$ with $h=0.7$ (${\rm H}_0 =
100 \hinv \kmsmpc$), and $M$ is the mass within a sphere
encompassing $200 
\rho_c(z)$ in units of $\massunits$.  Note the difference in Hubble
constant conventions for luminosity and mass.  We assume a concordance
cosmology with a lower power spectrum normalization that is consistent
with WMAP3 \citep{spergel:06}: $\{\Omega_m, \Omega_{\Lambda},
\sigate\} =\{0.3,0.7,0.8\}$.

First, we discuss mass selection in the presence of scatter and flux
limits and the consequences for \xray\ cluster
surveys (\S2).  We develop formalism in \S3 for studies of
correlated observable properties by introducing intrinsic covariance
within a joint distribution of luminosity and temperature.  Finally,
we address the role of covariance in the study of scaling relation
evolution. 

%%%%%%%%%%%%%%%%%%%%%%%%%%%%%%%%%%%%%%%%%%%%%%%%%%%%%%%%%%%%%%%%%%%%%
\section{L-M and Mass Selection} \label{sec:lm} %%%%%%%%%%%%%%%%%%%%%%%
%%%%%%%%%%%%%%%%%%%%%%%%%%%%%%%%%%%%%%%%%%%%%%%%%%%%%%%%%%%%%%%%%%%%%

We introduce the 
following notation for the logarithms of mass, luminosity, and temperature,
respectively: $\mpar\equiv \lnM$, $\lpar\equiv \lnL$, and $t\equiv
\lnT$.  We assume that the soft-band \xray\ luminosity of halos
follows a log-normal distribution with dispersion, $\sigl$,
about a mean that varies with mass and epoch as 
\beq
\bar{\lpar}=\lpar_{15,0} + p\,\mpar + s\,{\rm ln}[E^2(z)]. \label{eq:LM}  
\eeq
The normalization, $\lpar_{15,0}$, defines the present-epoch geometric
mean luminosity at the scale of $10^{15}\,h^{-1} \msol$;  $p$ is the
mass-scaling 
slope; and $E(z) \equiv \sqrt{\Omega_m(1+z)^3 +  \Omega_{\Lambda}}$ represents
the Hubble parameter evolution for a flat metric.  Our default model
uses $s=1$, appropriate for self-similar evolution
in the soft \xray\ band.  The scatter $\sigl$ is assumed to be mass-
and redshift-independent.  In what follows, we employ values $p \se
1.6$, $\sigl \se 0.59$, and \lnlnorm $= 1.81$, derived by matching
REFLEX counts to  predicted cluster counts to the Jenkins mass
function \citep{stanek:06}.    

%%%%%%%%%%%%%%%%%%%%%%%%%%%%%%%%%%%%%%%%%%%%%%
\FIGwindow %%%%%% PLOT: Window Function %%%%%%
%%%%%%%%%%%%%%%%%%%%%%%%%%%%%%%%%%%%%%%%%%%%%%

A sample's flux threshold, $\fluxlim$, sets a redshift-dependent luminosity 
limit, $\lmin(z)= {\rm log}[\,4\pi d_L^2(z)$  $\fluxlim/K(z,T)]$,
where the K-correction, $K(z,T)$, is detailed in the Appendix.
Temperatures are discussed in \S3.1 below.  
For non-negligible scatter, $\sigl$, this sharp luminosity-selection 
maps to a smooth selection, or window, function in
mass    
\beq
W(\mpar,z)= 
\int_{\lmin}^{\infty} \rmd \lpar\, \probdist(\lpar|{z},\mpar) =
\sqrt{\frac{\pi}{2}}{\rm erfc}(\ldiff), \label{eq:windfunc} 
\eeq
where $\probdist(\lpar|{z},\mpar)$ is a Gaussian distribution with
mean given by equation~(\ref{eq:LM}) and constant scatter $\sigl$,  
and $\ldiff\equiv
\left[\lmin(z,T)-\lbar(\mpar,z)\right]/\sqrt{2}\sigl$.   

Fig.~\ref{fig:windz} displays the window functions of current 
and next-generation \xray\ surveys for mass scales of the Coma-cluster 
($10^{15} h^{-1} M_{\odot}$) and for halos near the limits of modern
Sunyaev-Zel'dovich surveys  \citep[$10^{14} h^{-1}
M_\odot$,][]{holder:01}.  Dashed lines show the 
cosmic cumulative fraction, $N(<z)/N_{tot}$, of 
halos within redshift, $z$.  For our chosen normalization $\sigma_8 =
0.8$, most \mhiscale\ halos lie within $z
\se 1$, while the \mloscale\ population continues growing to $z\se 2$.
The next-generation window function allows detection of $100\%$ of the
\mhiscale\ halos in the sky.  The decline in the window function at $z
\sim 0.5$ permits detection of $\sim 56\%$ of the universal
population at \mloscale.  By contrast, the shallow flux limits of
current surveys detects only $34\%$ and $ 0.06\%$ of these totals. 

Such incomplete mass sampling leads to a Malmquist bias that brightens the \lm\
relation relative to the intrinsic (mass-limited) population
\citep{stanek:06}.  The log-mean luminosity of halos in a flux-limited
sample is  
\beq
\langle \lpar(\mpar)\rangle = \frac{\int
dV(z)\,n(\mpar,z)\, \left[  \lbar(\mpar,z)W(z|\mpar)+ \sigl
  e^{-\ldiff^2}\right]} {\int dN_f(\mpar)}, 
\label{eq:lmmom}
\eeq
where $\rmd N_f(\mpar)\,= \rmd V(z) n(\mpar,z)\,W(\mpar,z)$, represents
the differential number of halos--within a comoving volume element,
$dV(z)$--whose luminosities satisfy the threshold, $\lmin(z,T)$.  
The second term manifests the bias such that when $\psi_{min}$ becomes
negligible and $\rm{W}$ remains appreciable (\ie when the median
luminosity is near the flux limit), the geometric mean is enhanced by
an amount comparable to the scatter, $\sigl$.

Fig.~\ref{fig:lmana} compares the \lm\ relation for
shallow and deep flux-limited surveys to the underlying, mass-limited
relation.  To discern between the effects of survey selection 
and redshift evolution, we plot the \emph{de-evolved} geometric mean,
$\langle L/E^{2s}\rangle$, using the model value, $s \se 1$.  Samples
with the shallow flux limit show a gradual deviation from a 
power law at the high-mass end.  At extremely high masses--above $3 \times
10^{15} \hinv\msol$--the bias disappears as the sample becomes
mass-complete.  For the deeper flux limit, surveys become mass-complete
above $2 \times 10^{14} \hinv\msol$ (see Fig.~\ref{fig:windz}), and
the resultant \lm\ relation is unbiased above this mass scale.  

The deep sample probes to higher redshift than the shallow sample.
But simply comparing the   
mean luminosity at fixed mass between these samples, without taking
into account the evolving Malmquist bias, would produce an incorrect
conclusion about luminosity evolution.    

%%%%%%%%%%%%%%%%%%%%%%%%%%%%%%%%%%%%%%%
\FIGlmanalytic %%%%% PLOT: <L(M)> %%%%%
%%%%%%%%%%%%%%%%%%%%%%%%%%%%%%%%%%%%%%%

%%%%%%%%%%%%%%%%%%%%%%%%%%%%%%%%%%%%%%%%%%%%%%%%%%%%%%%%%%%%%%%%%%%%%
\section{\lt\ and Correlation} \label{sec:lt} %%%%%%%%%%%%%%%%%%%%%%%
%%%%%%%%%%%%%%%%%%%%%%%%%%%%%%%%%%%%%%%%%%%%%%%%%%%%%%%%%%%%%%%%%%%%%

Since accurate total masses are difficult to obtain, existing studies
have used \xray\ temperature as a low-scatter mass proxy \citep{emn:96}.  
Several investigations of \xray\ evolution have compared 
\lt\ relations among  low- and high- redshift populations.
Early work by \cite{mushotzky:97}, \cite{fairley:00}, and \cite{novicki:02}
find trends that are consistent with no evolution ($s=0$) 
to $z\lta 0.8$.  A recent study by \cite{maughan:06} compares the
WARPS \citep{scharf:97} subsample with data from  \cite{arnaud:99}
and \cite{vikhlinin:02} 
to probe conventional and modified forms of
self-similar evolution.  They find consistency among the \lt\ slopes
of differing redshift populations to be consistent with each other,
but slightly \emph{steeper} 
than the self-similar model.  In a sample of 10 clusters at $z >
0.4$, \cite{kotov:05} measure evolution somewhat steeper than, but consistent 
with, self-similarity.  \cite{lumb:04} report near-self-similar
evolution, and acknowledge the possibility of a flux-limit bias in
their eight-cluster sample at $0.45 < z < 0.62$.        

\cite{balogh:06} compare observations with analytic models of the ICM,
finding that the degree of evolution in the scaling relations depends
strongly on that of the \lm\ scatter.  \cite{stanek:06} determine the
redshift-independent $\sigma_{\mpar|\lpar}$ in the presence of
a flux-limit; they also note that simulations suggest a weak intrinsic
correlation of the deviations in luminosity and temperature at fixed
mass. 

We next incorporate such covariance into scaling relation
analysis.  After computing observables under a multi-variate Gaussian
model, we add an explicit demonstration of the model using halos
from Hubble Volume (HV) sky survey catalogues \citep{evrard:02}.  Our
aim is to demonstrate the interplay between covariance and scaling
relation parameters for flux-limited samples.

%%%%%%%%%%%%%%%%%%%%%%%%%%%%%%%%%%%%%%%%%%%%%%%%%%%%%%%%%%%%%%%%%%%%%
\subsection{Joint Luminosity-Temperature Distribution} %%%%%%%%%%%%%
\label{sec:lt_covar} %%%%%%%%%%%%%%%%%%%%%%%%%%%%%%%%%%%%%%%%%%%%%%%%
%%%%%%%%%%%%%%%%%%%%%%%%%%%%%%%%%%%%%%%%%%%%%%%%%%%%%%%%%%%%%%%%%%%%%

To calculate the geometric mean and dispersion of the \lt\ relation,
we assume that the log-mean temperature of the intracluster medium
follows virial scaling behavior, 
\beq
\bar{t} =  t_{15} + \frac{2}{3} \mpar +\frac{1}{3}\ln[E^2(z)], 
\label{eq:TM}
\eeq
with normalization $t_{15} \se {\rm
  ln}(kT_{15})$ and $kT_{15} \se 6.8 \keV$. 
We further assume that there is constant log-normal scatter about the
mean with $\sigma_{t}=0.10$, equivalent to a $15\%$ dispersion in mass
at fixed temperature. 

%%%%%%%%%%%%%%%%%%%%%%%%%%%%%%%%%%%%%%%%%%%%%%%%%%%%%%%%%%%%%%%%%
\FIGltranalyticsim %%%%%% PLOT: LT_ana, LT_sim; sigmaT_ana %%%%%%
%%%%%%%%%%%%%%%%%%%%%%%%%%%%%%%%%%%%%%%%%%%%%%%%%%%%%%%%%%%%%%%%%

Using the normalized deviations,
$\delta_{\lpar} \equiv 
\left(\lpar-\lbar\,\right)/\sigl$ and $\delta_t \equiv
\left(t-\bar{t}\,\right)/\sigma_t$, we form the
joint log-normal distribution of luminosity and temperature at a given
mass and epoch,      
\beq
\probdist(\lpar,t|\mpar,z)=\frac{1}{2\,\pi\, R} {\rm
  e}^{-\mbox{\boldmath$\delta$}^TC^{-1}\mbox{\boldmath$\delta$}}, 
\qquad
\mbox{\boldmath$\delta$} \equiv 
\left[   \begin{array}{c}
    \delta_{\lpar}\\ \delta_{t}
  \end{array}  \right], 
\label{eq:jointdist} 
\eeq
where $C_{ij} \equiv \langle \delta_i\delta_j \rangle$ is the
correlation matrix with off-diagonal elements equal to the correlation
coefficient, $r \equiv \langle\delta_{\lpar}\delta_t\rangle$, and  $R
\equiv \sqrt{1-r^2}$.  The $m^{\rm th}$ moment of the distribution at
fixed temperature is 
\beqnarr
\langle \lpar^m(t) \rangle \hspace{-5pt} \aqual \hspace{-5pt} \frac{\int \rmd V(z) \int \rmd \mpar\,
  n(\mpar,z)\, \tdist \bigL^{(m)}(t|\mpar,z)}{\int \rmd N_f(t)}  
\label{eq:ymoment}\\
\bigL^{(m)}(t|\mpar,z) \aqual \int \rmd\lpar\, \lpar^m\,
\probdist(\lpar|z,\mpar,t),
\eeqnarr
where $\rmd N_f(t) = \rmd V(z){\rm
  d}\mpar\,n(\mpar,z)W(\mpar,z)\tdist$.  The first moment (\ie\ the
geometric mean luminosity) is  
\beq
\Lone(t|\mpar,z) = W(\mpar,z)\, \ltrprox  + \sigl R\,  \rm
e^{-\phi_{min}^2} 
\label{eq:ytmean}, 
\eeq
where $\ltrprox \equiv \lbar + r\,\sigl\,\delta_t$ and 
% $\phit \se \left(\delta_{\lpar,min}-r\,\delta_t\right) R/\sqrt{2}$. 
$\phit \se \left(\lpar- \ltrprox\right)/ \sqrt{2}R\sigl$. 
At a given mass, non-zero correlation effectively shifts the mean
luminosity, either enhancing or suppressing the probability that a
halo of given mass survives the flux cut.  This has important
implications in the counts and mean masses discussed below.  In the
no-correlation limit ($r \se 0$), equations~(\ref{eq:ymoment}) and 
(\ref{eq:ytmean}) reduce to equation~(\ref{eq:lmmom}), but convolved
with a temperature-selection filter.     

From the \lt\ distribution's second moment,
\beqnarr
\Lsecondmom (t|\mpar,z) \aqual  W(\mpar,z)\left(\ltrprox^2 - R^2\sigl^2\right)
\nonumber \\  
& & + e^{-\phit^2}\left(2\,R\,\sigl\,\ltrprox  +
\sqrt{2}R^2\sigl^2\,\phit\right)\label{eq:ysecond},
\eeqnarr
we compute the observable variance,  $\sigl^2=\langle
\lpar^2\rangle - \langle \lpar \rangle^2$.  

In Fig.~\ref{fig:ltranasim}, we show the expected \lt\ relation mean 
and dispersion for REFLEX flux-limited clusters assuming three degrees
of \lt\ correlation, $r\in (-0.7,0.0,0.7)$.  Points in the figure
represent explicit realizations of the model derived from HV sky
survey samples of massive halos above $5\times\,10^{13} \hinv\msol$.
Using redshifts and masses of an all-sky survey, we assign
luminosities and temperatures via equations (\ref{eq:LM}) and
(\ref{eq:TM}), using two independent random Gaussian deviates, $g_1$
and $g_2$: $g_1$ controls the luminosity, and $g_t\ =\ r\, g_1 +  R\,
g_2$ sets the temperature. We then apply a flux limit including
K-correction terms.  

Similar to the flux-limited \lm\ relation
(Fig.~\ref{fig:lmana}), the \lt\, relation deviates from a pure power-law, although
this deviation is extremely weak for values, $r \gta 0.5$.  
The scatter depends strongly on the correlation coefficient, with 
magnitude varying from $0.8$ to $0.4$ as $r$ varies from
$-0.7$ to $0.7$.  Anti-correlation causes scatter orthogonal to the
mean input relation, enhancing the magnitude of the observable
dispersion.  Positive correlation reduces this dispersion.   

%\cite{markevitch:98} fit a power-law to a population of low-redshift
%clusters, uncorrected for cool cores.  The resulting prediction of the
%soft-band \xray\ luminosity, $L_{X,\left[0.1,2.4 \right]}=1.6\pm0.2\times\,
%\lumunits $, for a $4 \kev$ cluster lies slightly nearer to our models
%with {\it non-negative} correlation.

%Gasdynamic simulations show weakly positive correlation, $r \sims
%0.2$ \citep{stanek:06}, but the sensitivity of this result to physical
%treatment of the gas is unclear.  
% Comparison of the model predictions to 
% the growing body of \lt\ observational data should yield constraints on
% the correlation, but we defer that exercise to future work.  We turn
% next to observable features at high 
% redshift that next generation \xray\ samples will provide.  

%%%%%%%%%%%%%%%%%%%%%%%%%%%%%%%%%%%%%%%%%%%%%%%%%%%%%%%%%%%%%%%%%%%
 \FIGscatterevolution %%%%% PLOT: LZ|T_ana, multiple sig_lnL; %%%%%
                      %%%%%       LZ|T_ana, multiple r        %%%%% 
%%%%%%%%%%%%%%%%%%%%%%%%%%%%%%%%%%%%%%%%%%%%%%%%%%%%%%%%%%%%%%%%%%% 

%%%%%%%%%%%%%%%%%%%%%%%%%%%%%%%%%%%%%%%%%%%%%%%%%%%%%%%%%%%%%%%%%%%%%
\subsection{Evolution Diagnostics}\label{sec:scat_evo}%%%%%%%%%%%%%%
%%%%%%%%%%%%%%%%%%%%%%%%%%%%%%%%%%%%%%%%%%%%%%%%%%%%%%%%%%%%%%%%%%%%%

For mass-complete samples, the mean \lt\ relation will behave as 
$L \spropto T^{3p/2} \rho_c(z)^{s-p/2}$.  
For a flux-limited sample, the geometric mean soft-band luminosity at fixed
temperature is 
\beq
\langle \lpar(z|t)\rangle = \frac{\int \rmd\mpar\, n(z,\mpar)\int
  \rmd\lpar\, \lpar\, \probdist(\lpar|z,\mpar,t)} {\int \rmd n_f(z|t)},\label{eq:lztana}
\eeq
where $\rmd n_f(z|t) = \rmd\mpar\,n(\mpar,z)W(\mpar,z)\tdist$.  
The luminosity in equation~(\ref{eq:lztana})
and the sky surface density, $(4 \pi)^{-1}{\rm  d}V(z)/{\rm d}z {\rm d
}n_f(z|t)$, 
are shown in Fig.~\ref{fig:ltranaz} for the case
of $kT \se 4\kev$ clusters selected with a next-generation
($10^{-14} \cgsflux $) flux limit.  

The left panels show models with different \lt\ correlation within
our default model with $(s,\sigl, {\rm  ln}L_{15,0}) \se
(1.0,0.59,1.81)$.  The survey is mass-complete at 4~keV out to $z
\sims 0.7$, independent 
of $r$, giving counts that are identical in this redshift
range.  The mean luminosities shift from the mass-limited
mean by an amount $r\,\sigl  \langle \delta_t \rangle$, where $\langle
\delta_t \rangle$ is the mean temperature deviation of the selected
sample.  For halos of fixed temperature, the steepness of the mass
function implies that $\langle \delta_t \rangle > 0$; there are more
low-mass halos to scatter upward into the temperature bin than
high-mass halos to scatter downward.  This effect either  
brightens or dims the mean luminosity at low redshifts, 
depending on the sign of the correlation, $r$.  At $z \gta 0.7$, the
differences in 4~keV mass selection among the models become
appreciable.  Compared to the $r \se 0$ case, this selection effect
drives the counts up and the mean masses down for
the $r \se 0.7$ model, and vice-versa for $r \se -0.7$.  The magnitude
of this mass shift is small, however, amounting to $ -3\%$ and $ +6\%$
at $z \se 1.5$, respectively.   

The right panels of Fig.~\ref{fig:ltranaz} show two zero correlation models
tuned to produce behavior similar to the $r \se \pm 0.7$ cases.  The 
parameter sets, $(s,\sigl, {\rm  ln}L_{15,0}) \se (1.10,0.35,1.95)$ and
$(0.95,0.72,1.65)$, are within the $95\%$ error contours that result from the
REFLEX analysis of \cite{stanek:06}.  The solid line reproduces our
default model, as in the left panels, where $r \se 0$.  The luminosity
moments and sky surface densities of these models lie within a few
percent of the respective covariance models shown in the left panels,
but the driving mechanisms are different.  The dimmer intercept and
larger scatter of the $s \se 0.95$ case force this model to become
mass-incomplete earlier than the default case.  The counts drop beyond
$z \gta 0.7$ and the mean luminosity rises as the Malmquist bias is
triggered.  The $s \se 1.10$ case, with a smaller scatter and higher
intercept, remains mass-complete to a higher redshift, leading to
larger counts and a smaller Malmquist bias.  

Unlike the $r \se \pm 0.7$ case, the mean halo mass selected by the 
4~keV constraint is not shifted at high 
redshift in these models.  When $r \se 0$, halos that scatter up or down into a fixed
temperature bin are equally likely to be culled by the sample flux limit.   
Mass estimates from weak lensing of stacked ensembles could potentially be a
useful discriminatory diagnostic, but systematic effects must be controlled at
the one-percent level at $z \ssimeq 1$.  The upside is that next-generation surveys will
produce large numbers of high-$z$ clusters.   According to  
Fig.~\ref{fig:ltranaz}, a 5000-square degree, deep-flux limit survey
will yield 445 clusters with temperatures between $3.8$ and 
$4.2 \kev$, and between $0.9$ and $1.1$ in redshift.

%%%%%%%%%%%%%%%%%%%%%%%%%%%%%%%%%%%%%%%%%%%%%%%%%%%%%%%%%%%%%%%%%%%%%
\section{Conclusions}\label{sec:conc} %%%%%%%%%%%%%%%%%%%%%%%%%%%%
%%%%%%%%%%%%%%%%%%%%%%%%%%%%%%%%%%%%%%%%%%%%%%%%%%%%%%%%%%%%%%%%%%%%%

The multivariate properties of massive halos selected by \xray\ flux-limited
surveys are sensitive to assumptions about the correlation between
bulk observable properties at fixed mass. Using a model
in which the \lm\ relation is tuned to match local REFLEX counts, we
show how the Malmquist bias in mean luminosity (which arises from
mass-incompleteness) disappears during the transition from shallow to
deep flux-limited samples. Although the underlying model is based on
power-law behavior, the geometric mean \lm\ relation from a
flux-limited sample will deviate from a pure power-law, with a kink at
the mass scale above which the survey becomes mass-complete.  For a
survey with a flux limit, $10^{-14} \cgsflux$, this feature is expected
to lie near $M \se 2 \times 10^{14} \hinv\msol$.    

Using a model with Gaussian covariance to describe how $L$ and $T$
jointly relate to mass and redshift, we compute luminosity moments and
sky counts of clusters at fixed \xray\ temperature.  The slope and
scatter of the \lt\ relation for bright, local samples is sensitive to
the covariance between these properties.  Placing limits on the
covariance from local samples will require prior information on the
intrinsic $L$ and $T$ variance at fixed mass, and such priors can be
obtained from external observations or from gas dynamic simulations.  

Finally, we address attempts to extract information on cluster
evolution from the behaviour of the \lt\ relation within a flux-limited
survey.  We show that non-zero \lt\ covariance affects counts and
luminosity moments as a function of redshift in a manner that is 
degenerate with redshift evolution at zero covariance.  

A deep \xray\ survey of clusters, by itself, is limited to the
information provided by counts and moments.  Combining such a sample
with optical and sub-millimeter observations offers the potential to break
the evolution-covariance degeneracies through lensing-mass estimates
and additional signatures that can be computed via extensions to the
approach introduced in this paper.

%%%%%%%%%%%%%%%%%%%%%%%%%%%%%%%%%%%%%%%%%%%%%%%%%%%%%%%%%%%%%%%%%%%%%
%\acknowledgments
\bigskip

We thank Kerby Shedden and Chris Mullis for valuable conversations.
This work is supported by the Michigan Space Grant Consortium, by NASA
grant NAG5-13378, by NSF ITR grant ACI-0121671 and by NASA through
Chandra Postdoctoral Fellowship grant number PF6-70042 awarded by the
Chandra X-ray Center, which is operated by the SAO for NASA under
contract NAS8-03060.  AEE acknowledges support from the Miller
Foundation for Basic Research in Science at University of California,
Berkeley.  Simulations used in this paper were carried out by the
Virgo Supercomputing Consortium using computers based at the Computing
Centre of the Max-Planck Society in Garching and at the Edinburgh
parallel Computing Centre. The data are publicly available at
http://www.mpa-garching.mpg.de/NumCos.

\appendix
%%%%%%%%%%%%%%%%%%%%%%%%%%%%%%%%%%%%%%%%%%%%%%%%%%%%%%%%%%%%%%%%%%%%%
\section{K-Correction} \label{app:kcorr} %%%%%%%%%%%%%%%%%%%%%%%%%%
%%%%%%%%%%%%%%%%%%%%%%%%%%%%%%%%%%%%%%%%%%%%%%%%%%%%%%%%%%%%%%%%%%%%%
To convert rest frame to observed soft-band ($0.5-2.0
\kev$) flux, we multiply the former by a correction factor obtained from
fitting the output of an 0.3 solar metallicity 
{\tt mekal} plasma model. Specifically, we use $\Lobs = K(z,kT)\,
\Lrest$ with 
\beqnarr
K(z,kT) \aqual 1+ z\, \left[\, K_1(kT) + K_2(kT)\, z\, \right]\label{eq:kcorr} \\
K_1(kT) \aqual \ka+ \lgT (\kb \kc\,\lgT ) \nonumber \\
K_2(kT) \aqual \kd+ \lgT (\ke \kf\,\lgT ), \nonumber 
\eeqnarr
and $kT$ in keV. The fit is accurate to a few percent within $z=2$.

%%%%%%%%%%%%%%%%%%%%%%%%%%%%%%%%%%%%%%%%%%%%%%%%%%%%%%%%%%%%%%%%%%%%%%

%%%%%%%%%%%%%%%%

% Bibliography %
%%%%%%%%%%%%%%%%
\bibliographystyle{mn2e}
\bibliography{nsre07_submit}

\end{document}